Characterizing the Radiation Dose to Measurement Accuracy Relationship across Multiple Metrics in Opportunistic Chest CT

Boyuan Li, BS[1], Carolyn C. Chang, MD[2], Jake J. Kim, BA[2], Jia Wang, PhD, DABR[3], Justin R Tse, MD[1], Natalie S. Lui, MD[2], Haiwei Henry Guo, MD, PhD[1], Adam S. Wang, PhD[1]

[1]Department of Radiology, Stanford University Medical Center, Stanford, California

[2]Department of Cardiothoracic Surgery, Stanford University Medical Center, Stanford, California

[3]Department of Environmental Health and Safety, Stanford University, Stanford, California

Corresponding author: Boyuan Li, email: lby@stanford.edu

## Abstract

**Objectives:** Opportunistic Computed Tomography (CT) enables secondary health assessment from clinically acquired scans without additional imaging, but its reliability in scans with reduced-dose conditions remains underexplored. This study aims to characterize the dose-performance relationship for opportunistic CT and disentangle the contributions of segmentation failure and dose-dependent Hounsfield unit (HU) bias to performance degradation.

**Materials and Methods:** Simulated low-dose CT images at 1-75% of full dose were generated from 50 paired full- and low-dose non-contrast chest CT scans (Mayo Low-dose CT dataset). An independent dataset of 22 paired photon-counting CT acquisitions at lung cancer screening (LCS) and chest x-ray-equivalent (CXR) dose levels provided parallel real-world evaluation. Quantitative metrics for hepatic steatosis, sarcopenia, osteoporosis, cardiomegaly, emphysema, and coronary artery calcification were obtained using a two-step pipeline of deep learning-based segmentation followed by quantitative metric extraction. Classification performance was evaluated against full-dose reference standards, with additional analyses isolating the contributions of segmentation error and HU bias. Agreement between dose levels was assessed using Bland-Altman and correlation analyses.

**Results:** Mean HU metrics (hepatic steatosis, sarcopenia) maintained classification accuracy to CXR-equivalent dose (3%), with performance primarily limited by HU bias rather than segmentation failure; bias correction improved accuracy from 88% (95% CI: 76–96%) to 96% (95% CI: 86–100%) for hepatic steatosis and from 84% (95% CI: 71–93%) to 90% (95% CI: 78–97%) for sarcopenia. Trabecular bone attenuation (osteoporosis) maintained 98% (95% CI: 89–100%) accuracy at LCS dose but degraded sharply at lower dose levels due to segmentation failure, with minimal benefit from bias correction. Volume metrics (cardiomegaly) achieved 94% (95% CI: 84–99%) accuracy at CXR-equivalent dose, reflecting strong segmentation performance. Threshold-based metrics (emphysema) required LCS dose for reliable classification; bias correction improved accuracy from 58% (95% CI: 43–72%) to 92% (95% CI: 81–98%). Coronary artery calcification scoring failed below 5% dose but reached 96% (95% CI: 86–100%) accuracy at LCS dose. These results demonstrate that dose robustness varies by disease metric, informing which dose levels specific metrics remain reliable. In both Mayo and photon-counting CT datasets, agreement analyses at clinically relevant dose levels (LCS or CXR-equivalent) demonstrated strong correlation ($R^2 > 0.89$) for all metrics except coronary artery calcification, which was limited by clustering of values near zero.

**Conclusions:** Opportunistic CT assessment is feasible at reduced dose levels though it becomes less robust at ultra-low doses below 3-10% dose. Distinct failure modes are caused by HU bias or segmentation failure and depend on the clinical task. Providers should be aware of these task-specific limitations when designing opportunistic screening programs.

## 1. Introduction

Computed tomography (CT) has become one of the most widely utilized imaging modalities in modern medicine, with examination rates in the United States nearly tripling over the past two decades and continuing to grow at 2-5% annually in recent years[1-3]. Clinical interpretation remains appropriately focused on the primary diagnostic question, though additional rich tissue information is underutilized. In parallel, deep learning methods for medical image analysis have advanced rapidly and achieve strong performance across a range of tasks, with over 700 FDA-cleared algorithms in radiology alone[4,5]. The convergence of these trends has given rise to opportunistic CT. In this paradigm, clinically acquired scans are repurposed to extract clinically meaningful information beyond their original diagnostic purpose, without additional imaging, cost, or radiation exposure[6,7]. By enabling secondary assessment of conditions such as osteoporosis, hepatic steatosis, and cardiovascular diseases from scans obtained for unrelated indications, opportunistic CT offers a pathway to convert otherwise underutilized imaging data into actionable assessments for population-level screening and preventive care [7-9].

In parallel, CT practice has undergone a sustained shift toward lower radiation dose per exam, motivated in part by concerns that CT radiation exposure may contribute to future cancer incidence[10]. Advances in detector technology, iterative and deep learning-based reconstruction, and dose-optimization strategies have enabled substantial dose reductions while maintaining diagnostic image quality, driving widespread adoption of low-dose CT (LDCT)[11-14]. LDCT is now routine for lung cancer screening, and its use is expected to grow as radiation stewardship becomes increasingly central to clinical practice[15,16]. As more CT examinations are acquired at reduced dose, the volume of LDCT data potentially available for secondary assessment expands accordingly. Moreover, while opportunistic CT research has focused predominantly on abdominal CT[6,17], chest CT accounts for approximately 20 million of the over 90 million CT examinations performed annually in the United States [10] and may offer complementary opportunistic assessments or serve as the primary source of secondary health information for patients without existing abdominal CT.

Among the many dimensions of CT protocol variability that can affect HU-based opportunistic assessments, including tube voltage, reconstruction method, and scanner hardware[18-20], radiation dose is of particular importance. LDCT utilization is expanding rapidly, and reduced dose is well known to increase image noise and degrade HU accuracy[21]. This directly affects opportunistic CT pipelines, which typically consist of two sequential steps, automated anatomical segmentation followed by quantitative metrics extraction, either or both of which may be compromised under reduced-dose conditions. Recent work has begun to evaluate how automated segmentation and quantification perform at reduced dose. One study evaluated deep learning segmentation models across a range of simulated dose levels, finding that TotalSegmentator maintains high accuracy even at 20% of standard dose[22]. Another demonstrated that automated bone mineral density measurements on LDCT can effectively identify patients at risk for osteoporosis in lung cancer screening populations[23]. Similar single-dose evaluations on LDCT have been reported for hepatic steatosis and sarcopenia[24,25]. However, these efforts have complementary limitations. Studies evaluating segmentation robustness have characterized technical performance across a range of dose levels but have not linked their findings to clinical endpoints. Conversely, studies demonstrating clinical utility have typically evaluated only a single dose level, leaving the relationship between dose and diagnostic performance uncharacterized. Furthermore, neither approach has disentangled the sources of error, whether degraded performance reflects segmentation failure or HU bias, or both. The dose-performance relationship for opportunistic CT, including whether robustness varies across metrics and whether degradation stems from segmentation failure or HU bias, remains uncharacterized.

In this study, we characterize the dose-performance relationship for opportunistic CT by evaluating automated segmentation and downstream quantitative metrics across a range of simulated dose levels. Using paired full- and low-dose CT data, we disentangle the contributions of segmentation failure and dose-dependent HU bias to classification performance degradation. We complement simulation-based findings with a parallel evaluation using real-world paired photon-counting CT acquisitions at different dose levels. This study focuses on chest CTs and associated opportunistic measurements commonly reported in literature. Our results demonstrate that robustness varies by disease metric, with implications for determining the dose levels at which specific opportunistic assessments remain reliable.

## 2. Materials and Methods

This study employed simulated low-dose CT (LDCT) chest scans derived from the Mayo dataset to evaluate dose-dependent performance of opportunistic CT assessment using established clinical criteria. In addition, an independent photon-counting CT (PCCT) dataset containing paired low-dose and ultra-low-dose chest acquisitions was analyzed to provide a parallel evaluation using real-world clinical data. Quantitative metrics were extracted using a two-step pipeline consisting of deep learning-based organ segmentation followed by quantitative metric extraction. Dose-dependent performance on the Mayo dataset was evaluated using classification metrics and agreement analyses, with agreement analyses additionally applied to the PCCT dataset.

### 2.1 Low-Dose CT Simulations (Mayo Dataset)

#### *2.1.1 Mayo Low-dose CT Dataset*

The Mayo low-dose CT dataset provided paired non-contrast chest CT scans consisting of acquired full-dose images and corresponding simulated low-dose images at 10% of the original dose. A total of fifty paired scans were used in this study, with full-dose images acquired at a mean CTDIvol of 6.4 mGy. These paired images were perfectly registered, enabling subsequent simulation of additional intermediate dose levels.

#### *2.1.2 Image-Domain Low-dose CT Simulation*

To simulate CT images at arbitrary dose levels, an image-domain noise insertion method based on quantum noise properties was used (Figure 1). Assuming that quantum noise variance is inversely proportional to radiation dose ($\sigma^2 \propto 1/Dose$), paired full- and low-dose CT scans were used to estimate dose-dependent noise characteristics. A noise-only image ($Im_{Noise}$) was computed by subtracting the full-dose image ($Im_{FD}$) from the corresponding low-dose image ($Im_{LD}$), which is modeled as a zero-mean Gaussian noise with variance approximately equal to $(1/D_{LD} - 1)\sigma_{FD}^2$, where $D_{LD} = 10\%$ in this case. Under this noise model, the low-dose image can be expressed as the sum of the full-dose image and an additive noise term:

$$Im_{LD} = Im_{FD} + Im_{Noise} \sim N(\mu,\ \sigma_{FD}^2 + (1/D_{LD} - 1)\sigma_{FD}^2),$$

$$\text{where } Im_{Noise} \sim N(0,\ (1/D_{LD} - 1)\sigma_{FD}^2).$$

Then, CT images at desired target dose level $D_{Target}$ were generated by linearly combining the full-dose image with a scaled version of the noise image:

$$Im_{Target} = Im_{FD} + a\, Im_{Noise}.$$

The scaling factor $a$ was selected to match the expected noise variance at the target dose level and was computed as:

$$a = \sqrt{(1/D_{Target} - 1)/(1/D_{LD} - 1)}.$$

Using this approach, simulated CT images were generated at target dose levels corresponding to 1%, 2%, 3%, 5%, 25%, 50%, and 75% of the full dose. The 25% dose level approximates lung cancer screening (LCS) conditions, and the 3% dose level approximates chest x-ray-equivalent (CXR) dose. Since the noise image is derived from realistically simulated low-dose data rather than synthetic Gaussian noise, it captures realistic noise texture and spatial correlations present in reduced-dose acquisitions, although it likely underestimates electronic noise at ultra-low dose.

**2.2 Photon-counting CT Data Acquisition**

An independent PCCT dataset of paired real-world acquisitions at different dose levels was analyzed alongside the simulation-based study, providing a parallel evaluation using clinical data. This dataset consisted of paired low-dose and ultra-low-dose non-contrast chest PCCT acquisitions from a single-center lung cancer screening study at Stanford Health Care[26], retrospectively analyzed for opportunistic quantitative assessment. The ultra-low-dose scan was performed as a research study under Stanford IRB approval and informed consent. A total of 22 participants underwent both dose levels in consecutive scans during the same imaging session, in which the low-dose scan was a clinically indicated study and the ultra-low-dose scan was performed as a research study. All scans were acquired on a NAEOTOM Alpha scanner (Siemens Healthineers, Forchheim, Germany). Although the primary clinical task of lung nodule assessment relies on bone reconstruction kernels, soft-tissue kernel (Br36) reconstructions at 1mm slice thickness available within the same dataset were de-identified and used exclusively for opportunistic quantitative analysis since these metrics do not rely on high spatial resolution. Low-dose scans were acquired using a standard lung cancer screening (LCS) protocol at 120 kVp, while ultra-low-dose scans targeted chest x-ray-equivalent dose using a 100 kVp spectrum with tin filtration.

### 2.3 Quantitative metric extraction from CT images

Quantitative metrics were derived from automated segmentation of CT images for opportunistic assessment. Organ segmentation was performed using a deep learning-based segmentation tool built on the nnU-Net framework (TotalSegmentator, v2.4.0)[27,28], which produces segmentation masks for 117 anatomical structures, of which a subset of relevant organs was used for quantitative assessment. From these segmentation masks, four classes of quantitative metrics were computed: (1) whole-organ mean HU, (2) organ interior mean HU, (3) organ volume, and (4) the percentage of voxels below a predefined HU threshold.

Whole-organ mean HU was computed for the liver using the TotalSegmentator liver mask, where low mean HU is a predictor of hepatic steatosis[29]. For the erector spinae muscles, mean HU values were computed separately for the left and right erector spinae masks and then combined using a voxel-count-weighted average, equivalent to computing the mean HU over a single bilateral mask. A low mean HU is indicative of sarcopenia due to lower muscle mass and increased fat content[30]. For osteoporosis assessment, organ interior mean HU was computed for the L1 vertebral body, using a trabecular region of interest defined by applying a three-dimensional 3 mm morphological erosion to the whole vertebra segmentation mask to exclude the outer cortical bone. Low trabecular bone attenuation reflects reduced bone mineral density and increased fracture risk, indicative of osteoporosis[7]. For the lung, percent of voxels below -950 HU was computed across all pulmonary lobe masks using a volume-weighted aggregation, equivalent to evaluating the metric over a whole-lung mask. An elevated percentage of low-attenuation lung voxels indicates emphysematous tissue destruction[31]. Cardiac size was quantified using total heart volume from the whole-heart segmentation mask, without subdivision into individual cardiac chambers, where increased total heart volume reflects cardiomegaly associated with adverse cardiovascular outcomes[32]. Coronary artery calcium (CAC) scoring was performed separately using a pretrained deep learning model that operates on ungated chest CTs (AI-CAC, v1.0.0)[33]. CAC burden quantifies calcified atherosclerotic plaque in coronary arteries, with higher scores indicating greater cardiovascular risk[34].

For each organ, classification thresholds were selected based on established criteria from prior studies in consultation with clinical collaborators. The corresponding clinical conditions, metric definitions, and thresholds are summarized in Table 1. These thresholds correspond to mild-to-moderate disease severity, which is most relevant for opportunistic CT assessment focused on early detection.

To correct for dose-dependent HU bias in attenuation-based metrics, an organ-specific bias correction was applied. This reflects a calibration step that could be implemented in clinical deployment to account for systematic dose-dependent HU offsets. For each organ and dose level, the mean HU was computed across all subjects, and the difference from the corresponding full-dose mean was used to adjust low-dose HU measurements prior to applying classification thresholds.

**2.4 Statistical Analysis**

Statistical analyses were designed according to the structure of each dataset. For the Mayo LDCT dataset, analyses included dose-dependent classification performance evaluation across simulated dose levels and agreement comparisons between low- and full-dose images. For the PCCT dataset, analyses focused on agreement comparisons between paired low- and ultra-low-dose acquisitions.

For classification-based analyses in the Mayo LDCT dataset, quantitative metrics derived from the full-dose images were treated as the reference standard. For each metric, classification accuracy, sensitivity, and precision were computed by comparing low-dose classifications against the corresponding full-dose classifications using the thresholds summarized in Table 1. Segmentation performance was quantified using the Dice similarity coefficient, computed between low- and full-dose segmentation masks. Cases with missing metrics due to segmentation failure were treated as incorrect classifications when computing accuracy.

For HU-based metrics derived from organ segmentations, classification errors arise from two primary sources: segmentation error and dose-dependent HU bias. To isolate the relative contributions of these error sources, four analysis configurations were evaluated: (1) low-dose images with low-dose segmentation masks (baseline), (2) low-dose images with full-dose segmentation masks to isolate HU bias, (3) full-dose images with low-dose segmentation masks to isolate segmentation error, and (4) low-dose images with low-dose segmentation masks and bias correction. Classification performance under each configuration was evaluated using the same reference standards and metrics described above. Ninety-five percent confidence intervals for accuracy were computed using the Clopper-Pearson exact method.

Agreement between quantitative metrics derived from paired scans was assessed using correlation and Bland-Altman analyses. For the Mayo LDCT dataset, agreement was evaluated between full-dose images and selected simulated dose levels corresponding to LCS dose (25%) or CXR-equivalent dose (3%), depending on the metric evaluated. For the PCCT dataset, agreement was evaluated between LCS-dose and CXR-dose acquisitions.

## 3.Results

### 3.1 Mayo Low-dose CT Dataset

#### *3.1.1 Segmentation Robustness Across Dose Levels*

In Figure 2, example images from one subject demonstrate organ segmentations across axial, sagittal, and coronal views at full and ultra-low dose. Despite increased image noise from substantial dose reduction, segmentation remains consistent.

Segmentation robustness across simulated dose levels is summarized in Figure 3 using Dice similarity coefficient between low- and full-dose segmentation masks. Dice scores decreased monotonically with decreasing dose for all evaluated organs. Large soft-tissue organs such as liver and lung maintained high Dice scores at low dose levels, while smaller structures such as the L1 vertebral body exhibited reduced agreement. An inset for the L1 vertebral body illustrates reduced metric availability at lower dose levels due to segmentation failure. Metric availability for other organs remained near complete and is therefore not shown. Dice analysis was not applicable for coronary artery calcium, which was computed using a separate deep learning model.

*3.1.2 Classification Performance Across Dose Levels*

For large soft-tissue mean HU-based metrics, classification accuracy was maintained at ultra-low dose levels with modest accuracy degradation attributed to HU bias. As shown in Figure 4, hepatic steatosis and sarcopenia classification remained accurate down to 3% dose, with accuracy improving after bias correction from 88% (95% CI: 76-96%) to 96% (95% CI: 86-100%) for hepatic steatosis and from 84% (95% CI: 71-93%) to 90% (95% CI: 78-97%) for sarcopenia. In contrast, threshold-based metrics showed substantial accuracy degradation at low dose, strongly influenced by HU bias. Lung emphysema classification accuracy improved substantially under the bias-corrected configuration, from 58% (95% CI: 43-72%) to 92% (95% CI: 81-98%) accurate at 25% dose. For all three metrics, the segmentation-isolated (full dose image with low dose segmentation mask to only show segmentation error) configuration maintained high accuracy across dose levels, consistent with the improvements observed after bias correction, indicating that HU bias was the primary contributor to performance degradation.

Osteoporosis classification based on the L1 trabecular interior remained 98% (95% CI: 89-100%) accurate down to 25% dose, with accuracy degrading sharply at lower dose levels and minimal benefit from bias correction. This pattern suggests that segmentation failure, rather than HU bias, is the primary contributor to performance degradation. Consistent with this observation, the HU-bias-isolated (low dose image with full dose mask to only show HU error) maintained relatively high accuracy even at ultra-low dose.

Volume-based metrics for large soft-tissue organs also remained stable at ultra-low dose since they only depend on segmentation, with cardiomegaly classification achieving 94% (95% CI: 84-99%) accuracy down to 3% dose. For coronary artery calcification, the AI-CAC model failed to generate outputs at dose levels below 5%, but at 25% dose classification accuracy reached 96% (95% CI: 86-100%). This behavior indicates reduced robustness of the AI-CAC model at ultra-low dose, while still suggesting potential utility at low dose levels where the model produces valid outputs.

Classification sensitivity and precision trends are shown in Figure 5 and Figure 6, respectively. For large soft-tissue organs such as the liver and heart, classification sensitivity remained high across dose levels, indicating that dose-dependent performance was primarily limited by precision. For hepatic steatosis, bias correction substantially improved precision at low dose while preserving sensitivity. A similar false-positive-dominated pattern was observed for lung emphysema: sensitivity remained near the ceiling across dose levels, whereas precision was strongly affected by HU bias, and bias correction substantially improved precision, consistent with the accuracy trends. In contrast, sarcopenia classification exhibited a modest sensitivity-precision tradeoff at the lowest dose levels: while bias correction improved precision, a small decrease in sensitivity was observed at ultra-low dose. For coronary artery calcification, classification errors reflected a combination of false positives and false negatives at low dose, with false positives contributing slightly more to overall performance degradation.

Osteoporosis classification exhibited distinct sensitivity-precision behavior compared to other metrics. In the HU-bias-isolated configuration, sensitivity remained near the ceiling across dose levels, indicating that when a trabecular region was available, classification performance was primarily limited by precision, similar to other mean HU metrics. In the segmentation-isolated configuration, sensitivity trends were identical to baseline, consistent with the accuracy findings. With bias correction, osteoporosis classification precision improved but sensitivity decreased at lower dose levels. This pattern suggests that bias correction, while effective for metrics where HU bias is the primary error source, can produce unexpected behavior when segmentation failure dominates. Precision estimates are not directly comparable across configurations because missing values are excluded from precision calculations by definition. Reliable precision estimates were therefore obtained only for the HU-bias-isolated configuration.

### *3.1.3 Agreement Analysis at Clinically Relevant Dose Levels*

Agreement between low-dose and full-dose quantitative metrics in the Mayo LDCT dataset is shown in Figure 7 using Bland-Altman and correlation analyses. For each disease metric, results are presented at the lowest clinically relevant dose level (3% dose for CXR-equivalent dose and 25% for LCS-equivalent dose) achieving greater than 90% classification accuracy. Bland-Altman plots report low-dose minus full dose differences with corresponding bias and ±1.96 standard deviation limits, while correlation plots summarize linear agreement using linear regression and $R^2$ values.

For metrics previously identified as HU-error dominant, including hepatic steatosis at CXR-dose and lung emphysema at LCS-dose, agreement analyses showed a consistent low-dose bias with preserved proportionality. Sarcopenia metrics at CXR-equivalent dose showed systematic bias similar in magnitude to liver, but with wider limits of agreement and reduced $R^2$ due to outliers. Cardiomegaly metrics at CXR-equivalent dose showed good agreement, reflecting robust segmentation performance. Osteoporosis metrics at LCS dose demonstrated good agreement with limited bias and variability. For coronary artery calcification, values clustered near zero, resulting in reduced correlation and limiting the interpretability of correlation-based agreement.

### 3.2 PCCT Dataset

#### *3.2.1 Segmentation and Agreement Analysis between CXR-equivalent and LCS-dose PCCT*

In the PCCT dataset, the mean effective dose was 1.29±0.39 mSv (2.65±0.78 mGy) for LCS-dose acquisitions and 0.13±0.03 mSv (0.27±0.06 mGy) for CXR-equivalent dose acquisitions, representing a 90% dose reduction. Representative axial, sagittal, and coronal views are shown in Figure 8 for PCCT images acquired at LCS and CXR-equivalent dose levels with overlaid organ segmentations. At CXR-equivalent dose, PCCT images preserved sufficient soft-tissue contrast and anatomical delineation to support reliable organ segmentation.

Agreement between LCS and CXR-equivalent dose PCCT metrics using the soft-tissue reconstruction kernel is shown in Figure 9. Using the soft-tissue reconstruction kernel, PCCT-derived quantitative metrics demonstrated strong linear agreement across dose levels for mean HU metrics (hepatic steatosis, sarcopenia, and osteoporosis) as well as volumetric metrics (cardiomegaly). Bland–Altman analyses revealed consistent offsets between dose levels, indicating systematic bias rather than random disagreement. While this behavior suggests that bias correction may be feasible, the limited sample size and availability of only two dose levels precluded robust bias modeling in the present study. Among threshold-based metrics, emphysema showed strong agreement while coronary artery calcification exhibited greater dispersion, consistent with its sensitivity to image noise. Overall, these results demonstrate strong internal consistency between LCS and CXR-equivalent dose PCCT across quantitative metrics.

**4. Discussion**

In this study, we evaluated the dose-performance relationship of a representative set of opportunistic CT-derived quantitative metrics by disentangling segmentation-related errors from dose-dependent HU bias. Across metrics, clinically meaningful classification performance was maintained at substantially reduced dose levels, with the degree of robustness varying by metrics.

Differences in dose robustness across disease metrics reflected distinct failure modes in attenuation measurements rather than segmentation performance alone, which remains robust at low-dose conditions consistent with prior work[22]. For mean HU metrics, including those used for hepatic steatosis and sarcopenia, dose reduction primarily introduced systematic HU-bias, leading to shifts in measured attenuation even when segmentation remained stable. In contrast, threshold-based metrics, such as emphysema and coronary artery calcification, were more sensitive to increased variance at low dose, consistent with previously reported challenges in quantitative CT arising from noise amplification near fixed decision thresholds[35,36]. Notably, trabecular bone attenuation for osteoporosis assessment was influenced by both segmentation uncertainty and dose-dependent HU bias, consistent with the known sensitivity of trabecular bone attenuation measurements to ROI placement and size[37]. Together, these distinct failure mechanisms explain the heterogeneous dose response observed across disease metrics.

For PCCT, attenuation measurements remain dependent on acquisition parameters, and differences in tube potential and filtration (e.g., 100 kVp with tin filter versus 120 kVp) can introduce systematic shifts in HU[38]. Despite this, strong internal consistency was observed across CXR-equivalent and LCS-dose PCCT metrics, which indicate that opportunistic quantitative assessments remain reliable across low-dose regimes. Soft-tissue kernels were selected for this analysis because they produce smoother images with reduced noise, which is preferable for quantitative measurements. Studies with multicenter design and inclusion of patients outside of lung cancer screening would be needed to verify if our findings can be more broadly generalized.

Dose robustness in opportunistic CT should be interpreted relative to specific clinical imaging contexts. The dose ranges evaluated here align with established regimes such as CXR-equivalent imaging and lung cancer screening CT, representing realistic settings for opportunistic assessment. Recent advances in CT hardware and reconstruction have demonstrated the feasibility of thoracic imaging at CXR-equivalent dose levels, reinforcing the relevance of these regimes for emerging opportunistic screening applications[39,40]. Within these contexts, clinically meaningful classification could be maintained for multiple disease metrics despite dose-dependent HU bias and variance. However, the usable dose range differed across metrics, indicating that no single dose threshold applies across all opportunistic CT metrics and that dose selection must be tailored to the intended clinical task.

Prior opportunistic analyses of attenuation-based metrics using LDCT often compute mean HU metrics directly from LDCT, without accounting for the dose-dependent HU-bias and generally assuming sufficient HU stability at low dose[24,41,42]. When normalization is applied, it is typically performed using an internal reference within the LDCT scan, such as using spleen attenuation to compute muscle-to-spleen ratio for muscle density evaluation[25]. Threshold-based metrics, such as emphysema and coronary artery calcification, have also been studied in the opportunistic LDCT setting, where prior work has acknowledged increased noise sensitivity and applied noise-mitigation strategies[36,43,44], also without accounting for dose-dependent HU-bias. In this work, we explicitly evaluate the effect of dose-dependent HU bias and mitigate it by applying dose-dependent bias-correction. Moreover, prior opportunistic LDCT studies typically report performance at a single screening dose level defined by the corresponding dataset, despite substantial variation in LCS protocols across institutions. In this work, we characterize quantitative performance as a function of dose, enabling assessment of metric robustness and identification of dose levels at which performance degradation occurs.

This study has several limitations. First, the study cohort was a small, retrospective dataset. Second, the dataset did not incorporate clinical outcomes or diagnostic reference standards, as the focus was on relative dose robustness rather than absolute disease validation. Third, disease status was assessed using fixed binary classification thresholds, which simplifies interpretation but does not capture the full spectrum of disease severity or uncertainty near decision boundaries. The small sample size and binary thresholds limit statistical power, where minor non-monotonic patterns observed for some metrics likely reflect variability in threshold-boundary cases rather than systematic dose-dependent effects. Fourth, simulated low-dose images may not fully replicate the noise texture and artifacts present in physically acquired low-dose scans. Performance on physically acquired data may differ, particularly since electronic noise was not explicitly modeled in extrapolating the Mayo LCDT dataset. Additionally, results from PCCT may not generalize to conventional CT systems. Fifth, attenuation-based metrics relied on mean values within segmentation masks; exploratory erosion of these masks revealed intra-organ HU variability, indicating that sampling strategy can influence quantitative stability and warrants further investigation. Finally, deep learning-based segmentation was performed using models not explicitly optimized for ultra-low-dose imaging, and while segmentation robustness was generally preserved, the use of segmentation models optimized for low-dose imaging may further improve performance at reduced dose levels.

Future work will expand the availability of controlled multi-dose data, either through larger collections of paired full- and low-dose CT scans with registration or through validated LDCT simulation frameworks using full-dose images as input only. Clinically validated reference standards will also be incorporated for evaluation of opportunistic metrics, beyond internal consistency between dose levels. Future studies may also incorporate more granular disease stratification or continuous severity measures rather than a binary classification. Methodologically, integration of dose-aware segmentation or denoising approaches may further improve robustness of opportunistic quantitative assessments and expand the utility of existing LDCTs. While our focus was on chest CT, these methods should be evaluated in other exams such as abdominal CT. Lastly, extending opportunistic assessment to longitudinal settings represents an important direction. The low-dose nature of LDCT enables more frequent follow-up imaging, allowing evaluation of temporal changes at an individual and personalized level.

## 5. Conclusion

This study systematically characterized the dose-performance relationship for opportunistic CT. We demonstrate that clinically meaningful classification can be maintained at substantially reduced dose levels for multiple disease metrics. By disentangling segmentation failure from dose-dependent HU bias, we identified distinct failure modes across metrics: mean HU metrics were primarily limited by HU bias and benefited from bias correction, while threshold-based metrics showed greater sensitivity to noise, and trabecular bone assessment was dominated by segmentation failure at ultra-low dose. As LDCT utilization continues to expand, understanding these metric-specific dose tolerances will be essential for realizing the full potential of opportunistic CT in population-level screening and preventive care.

**Acknowledgement**: The authors thank Tie Liang, biostatistician, Stanford Radiology for reviewing statistical methodology.

**Disclosure**: None

## References

1. Berrington de González A, Mahesh M, Kim KP, et al. Projected cancer risks from computed tomographic scans performed in the United States in 2007. *Arch Intern Med*. Dec 14 2009;169(22):2071-7. doi:10.1001/archinternmed.2009.440
2. Smith-Bindman R, Kwan ML, Marlow EC, et al. Trends in Use of Medical Imaging in US Health Care Systems and in Ontario, Canada, 2000-2016. *JAMA*. Sep 03 2019;322(9):843-856. doi:10.1001/jama.2019.11456
3. Sarma A, Heilbrun ME, Conner KE, Stevens SM, Woller SC, Elliott CG. Radiation and chest CT scan examinations: what do we know? *Chest*. Sep 2012;142(3):750-760. doi:10.1378/chest.11-2863
4. Zhou SK, Greenspan H, Davatzikos C, et al. A review of deep learning in medical imaging: Imaging traits, technology trends, case studies with progress highlights, and future promises. *Proc IEEE Inst Electr Electron Eng*. May 2021;109(5):820-838. doi:10.1109/JPROC.2021.3054390
5. Sivakumar R, Lue B, Kundu S. FDA Approval of Artificial Intelligence and Machine Learning Devices in Radiology: A Systematic Review. *JAMA Netw Open*. Nov 03 2025;8(11):e2542338. doi:10.1001/jamanetworkopen.2025.42338
6. Pickhardt PJ. Value-added Opportunistic CT Screening: State of the Art. *Radiology*. May 2022;303(2):241-254. doi:10.1148/radiol.211561
7. Pickhardt PJ, Pooler BD, Lauder T, del Rio AM, Bruce RJ, Binkley N. Opportunistic screening for osteoporosis using abdominal computed tomography scans obtained for other indications. *Ann Intern Med*. Apr 16 2013;158(8):588-95. doi:10.7326/0003-4819-158-8-201304160-00003
8. Starekova J, Hernando D, Pickhardt PJ, Reeder SB. Quantification of Liver Fat Content with CT and MRI: State of the Art. *Radiology*. Nov 2021;301(2):250-262. doi:10.1148/radiol.2021204288
9. Pickhardt PJ, Graffy PM, Zea R, et al. Automated CT biomarkers for opportunistic prediction of future cardiovascular events and mortality in an asymptomatic screening population: a retrospective cohort study. *Lancet Digit Health*. Apr 2020;2(4):e192-e200. doi:10.1016/S2589-7500(20)30025-X
10. Smith-Bindman R, Chu PW, Azman Firdaus H, et al. Projected Lifetime Cancer Risks From Current Computed Tomography Imaging. *JAMA Intern Med*. Jun 01 2025;185(6):710-719. doi:10.1001/jamainternmed.2025.0505
11. Flohr T, Petersilka M, Henning A, Ulzheimer S, Ferda J, Schmidt B. Photon-counting CT review. *Phys Med*. Nov 2020;79:126-136. doi:10.1016/j.ejmp.2020.10.030
12. Hara AK, Paden RG, Silva AC, Kujak JL, Lawder HJ, Pavlicek W. Iterative reconstruction technique for reducing body radiation dose at CT: feasibility study. *AJR Am J Roentgenol*. Sep 2009;193(3):764-71. doi:10.2214/AJR.09.2397
13. Nagayama Y, Sakabe D, Goto M, et al. Deep Learning-based Reconstruction for Lower-Dose Pediatric CT: Technical Principles, Image Characteristics, and Clinical Implementations. *Radiographics*. 2021;41(7):1936-1953. doi:10.1148/rg.2021210105
14. Söderberg M, Gunnarsson M. Automatic exposure control in computed tomography--an evaluation of systems from different manufacturers. *Acta Radiol*. Jul 2010;51(6):625-34. doi:10.3109/02841851003698206
15. de Koning HJ, van der Aalst CM, de Jong PA, et al. Reduced Lung-Cancer Mortality with Volume CT Screening in a Randomized Trial. *N Engl J Med*. Feb 06 2020;382(6):503-513. doi:10.1056/NEJMoa1911793
16. Krist AH, Davidson KW, Mangione CM, et al. Screening for Lung Cancer: US Preventive Services Task Force Recommendation Statement. *JAMA*. Mar 09 2021;325(10):962-970. doi:10.1001/jama.2021.1117
17. Pickhardt PJ, Graffy PM, Perez AA, Lubner MG, Elton DC, Summers RM. Opportunistic Screening at Abdominal CT: Use of Automated Body Composition Biomarkers for Added Cardiometabolic Value. *Radiographics*. 2021;41(2):524-542. doi:10.1148/rg.2021200056

18. Zheng X, Al-Hayek Y, Cummins C, et al. Body size and tube voltage dependent corrections for Hounsfield Unit in medical X-ray computed tomography: theory and experiments. *Sci Rep*. Sep 24 2020;10(1):15696. doi:10.1038/s41598-020-72707-y
19. Szczykutowicz TP, Toia GV, Dhanantwari A, Nett B. A Review of Deep Learning CT Reconstruction: Concepts, Limitations, and Promise in Clinical Practice. *Current Radiology Reports*. 2022/09/01 2022;10(9):101-115. doi:10.1007/s40134-022-00399-5
20. Lamba R, McGahan JP, Corwin MT, et al. CT Hounsfield numbers of soft tissues on unenhanced abdominal CT scans: variability between two different manufacturers' MDCT scanners. *AJR Am J Roentgenol*. Nov 2014;203(5):1013-20. doi:10.2214/AJR.12.10037
21. Hsieh J. *Computed Tomography: Principles, Design, Artifacts, and Recent Advances*. 3rd ed. SPIE Press; 2015.
22. Tsanda A, Nickisch H, Wissel T, Klinder T, Knopp T, Grass M. Dose robustness of deep learning models for anatomic segmentation of computed tomography images. *J Med Imaging (Bellingham)*. Jul 2024;11(4):044005. doi:10.1117/1.JMI.11.4.044005
23. Kang WY, Yang Z, Park H, et al. Automated Opportunistic Osteoporosis Screening Using Low-Dose Chest CT among Individuals Undergoing Lung Cancer Screening in a Korean Population. *Diagnostics (Basel)*. Aug 16 2024;14(16)doi:10.3390/diagnostics14161789
24. Han S, Joo I, Park J, Jeon SK, Yoon SH. Clinical feasibility of fully automated three-dimensional liver segmentation in low-dose chest computed tomography (CT) for assessing hepatic steatosis: an alternative to abdominal CT. *Clin Radiol*. Nov 2025;90:107079. doi:10.1016/j.crad.2025.107079
25. Chen X, Gao X, Wang R, et al. Opportunistic muscle density assay during CT lung cancer screening for low muscle quality evaluation in older adults: a multicenter study. *Aging Clin Exp Res*. Feb 22 2025;37(1):41. doi:10.1007/s40520-025-02933-9
26. Chang CC, Li B, Kim J, et al. Chest x-ray dose photon-counting computed tomography enables comparable pulmonary nodule detection as low-dose computed tomography for lung cancer screening. *Western Thoracic Surgical Association Annual Meeting*. 2026
27. Wasserthal J, Breit HC, Meyer MT, et al. TotalSegmentator: Robust Segmentation of 104 Anatomic Structures in CT Images. *Radiol Artif Intell*. Sep 2023;5(5):e230024. doi:10.1148/ryai.230024
28. Isensee F, Jaeger PF, Kohl SAA, Petersen J, Maier-Hein KH. nnU-Net: a self-configuring method for deep learning-based biomedical image segmentation. *Nat Methods*. Feb 2021;18(2):203-211. doi:10.1038/s41592-020-01008-z
29. Zhang YN, Fowler KJ, Hamilton G, et al. Liver fat imaging-a clinical overview of ultrasound, CT, and MR imaging. *Br J Radiol*. Sep 2018;91(1089):20170959. doi:10.1259/bjr.20170959
30. Aubrey J, Esfandiari N, Baracos VE, et al. Measurement of skeletal muscle radiation attenuation and basis of its biological variation. *Acta Physiol (Oxf)*. Mar 2014;210(3):489-97. doi:10.1111/apha.12224
31. Johannessen A, Skorge TD, Bottai M, et al. Mortality by level of emphysema and airway wall thickness. *Am J Respir Crit Care Med*. Mar 15 2013;187(6):602-8. doi:10.1164/rccm.201209-1722OC
32. Sosnowski M, Parma Z, Syzdół M, et al. A Novel Concept of the "Standard Human" in the Assessment of Individual Total Heart Size: Lessons from Non-Contrast-Enhanced Cardiac CT Examinations. *Diagnostics (Basel)*. Jun 13 2025;15(12)doi:10.3390/diagnostics15121502
33. Hagopian R, Strebel T, Bernatz S, et al. AI Opportunistic Coronary Calcium Screening at Veterans Affairs Hospitals. *NEJM AI*. 2025;2(6):AIoa2400937. doi:doi:10.1056/AIoa2400937
34. Rumberger JA, Brundage BH, Rader DJ, Kondos G. Electron beam computed tomographic coronary calcium scanning: a review and guidelines for use in asymptomatic persons. *Mayo Clin Proc*. Mar 1999;74(3):243-52. doi:10.4065/74.3.243
35. Gierada DS, Pilgram TK, Whiting BR, et al. Comparison of standard- and low-radiation-dose CT for quantification of emphysema. *AJR Am J Roentgenol*. Jan 2007;188(1):42-7. doi:10.2214/AJR.05.1498
36. Jacobs PC, Isgum I, Gondrie MJ, et al. Coronary artery calcification scoring in low-dose ungated CT screening for lung cancer: interscan agreement. *AJR Am J Roentgenol*. May 2010;194(5):1244-9. doi:10.2214/AJR.09.3047

37. Sobecki JN, Krueger D, Pickhardt PJ, Binkley N, Anderson PA. Optimizing region of interest size and placement for clinical opportunistic CT trabecular bone Hounsfield unit measurements. *Osteoporos Int*. Sep 2025;36(9):1573-1581. doi:10.1007/s00198-025-07529-7
38. Yang Y, Qin L, Lin H, et al. Consistency of Monoenergetic Attenuation Measurements for a Clinical Dual-Source Photon-Counting Detector CT System Across Scanning Paradigms: A Phantom Study. *AJR Am J Roentgenol*. May 2024;222(5):e2330631. doi:10.2214/AJR.23.30631
39. Kroschke J, Kerber B, Eberhard M, Ensle F, Frauenfelder T, Jungblut L. Photon-Counting Chest CT at Radiography-Comparable Dose Levels: Impact on Opportunistic Visual and Semiautomated Coronary Calcium Quantification. *Invest Radiol*. Jan 01 2026;61(1):41-48. doi:10.1097/RLI.0000000000001199
40. Kerber B, Ensle F, Kroschke J, et al. The Effect of X-ray Dose Photon-Counting Detector Computed Tomography on Nodule Properties in a Lung Cancer Screening Cohort: A Prospective Study. *Invest Radiol*. Oct 01 2025;60(10):627-635. doi:10.1097/RLI.0000000000001174
41. Weiss J, Bernatz S, Johnson J, et al. Opportunistic assessment of steatotic liver disease in lung cancer screening eligible individuals. *J Intern Med*. Mar 2025;297(3):276-288. doi:10.1111/joim.20053
42. Cheng X, Zhao K, Zha X, et al. Opportunistic Screening Using Low-Dose CT and the Prevalence of Osteoporosis in China: A Nationwide, Multicenter Study. *J Bone Miner Res*. Mar 2021;36(3):427-435. doi:10.1002/jbmr.4187
43. Labaki WW, Xia M, Murray S, et al. Quantitative Emphysema on Low-Dose CT Imaging of the Chest and Risk of Lung Cancer and Airflow Obstruction: An Analysis of the National Lung Screening Trial. *Chest*. May 2021;159(5):1812-1820. doi:10.1016/j.chest.2020.12.004
44. Mulshine JL, Aldigé CR, Ambrose LF, et al. Emphysema Detection in the Course of Lung Cancer Screening: Optimizing a Rare Opportunity to Impact Population Health. *Ann Am Thorac Soc*. Apr 2023;20(4):499-503. doi:10.1513/AnnalsATS.202207-631PS

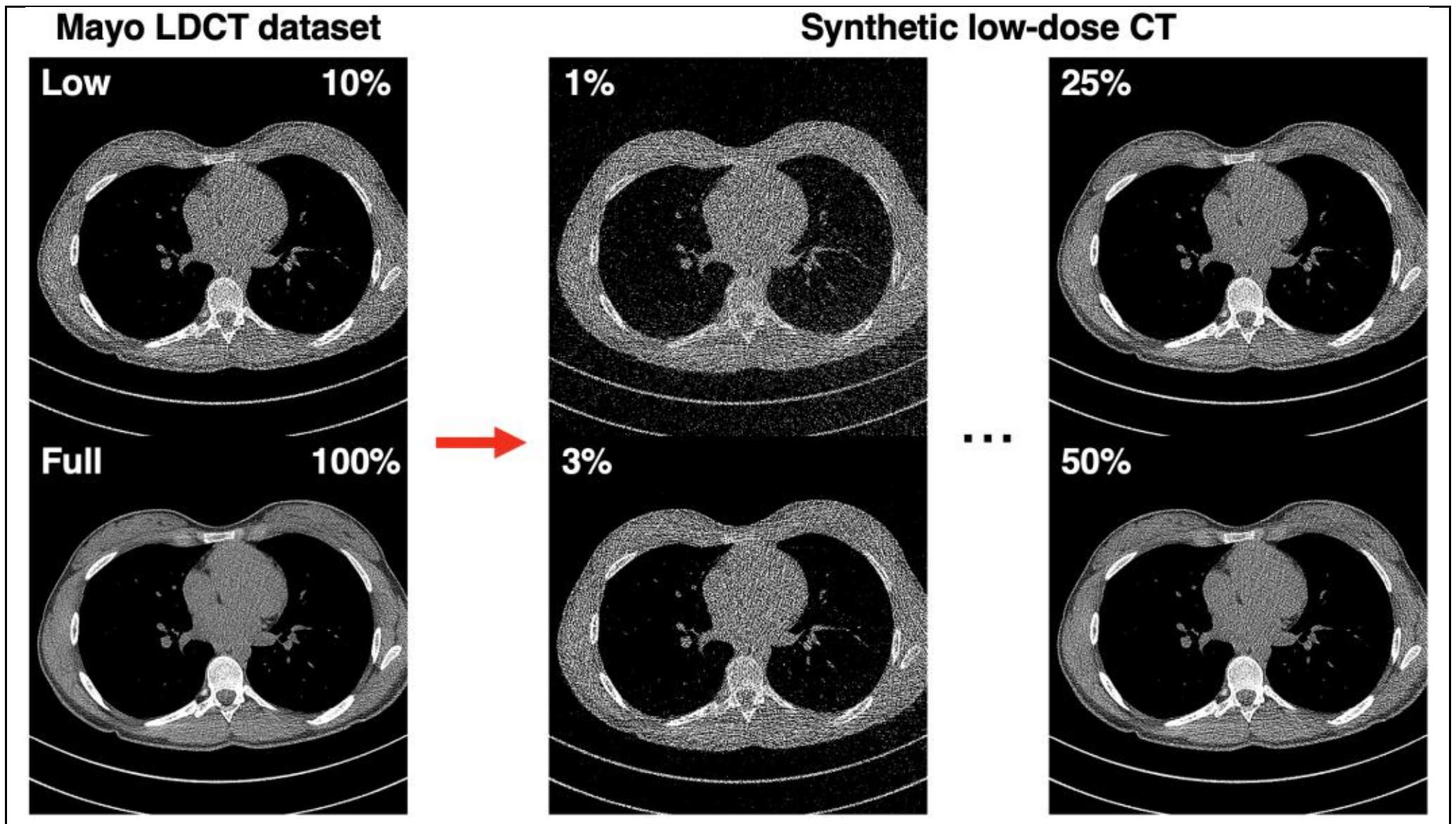

**Figure 1.** Image domain simulation of low-dose CT (LDCT) images from paired full- and low-dose images from the Mayo LDCT dataset. A representative axial slice at the mid-thorax is shown at both original and selected simulated dose levels (W/L: 400/40 HU).

HU, Hounsfield unit.

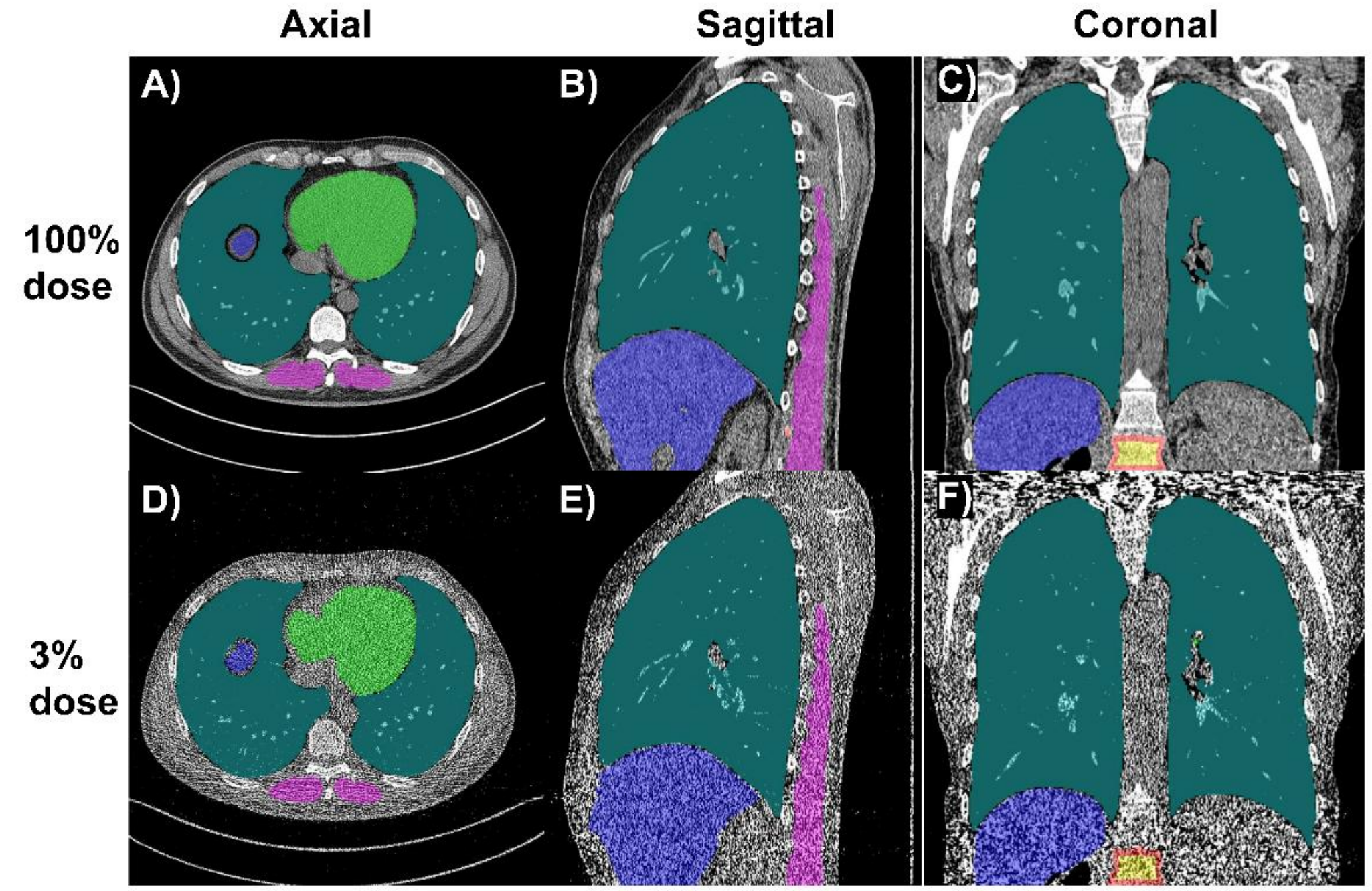

**Figure 2.** Representative axial, sagittal, and coronal views are shown for full-dose (100%; A-C) and low-dose (3%; D-F) CT images with overlaid organ segmentations obtained from the respective dose levels (Window/Level: 400/40 HU, liver (blue), lungs (teal), erector spinae (magenta), L1 vertebra (red; yellow indicates eroded mask for trabecular region), heart (green)).

HU, Hounsfield unit.

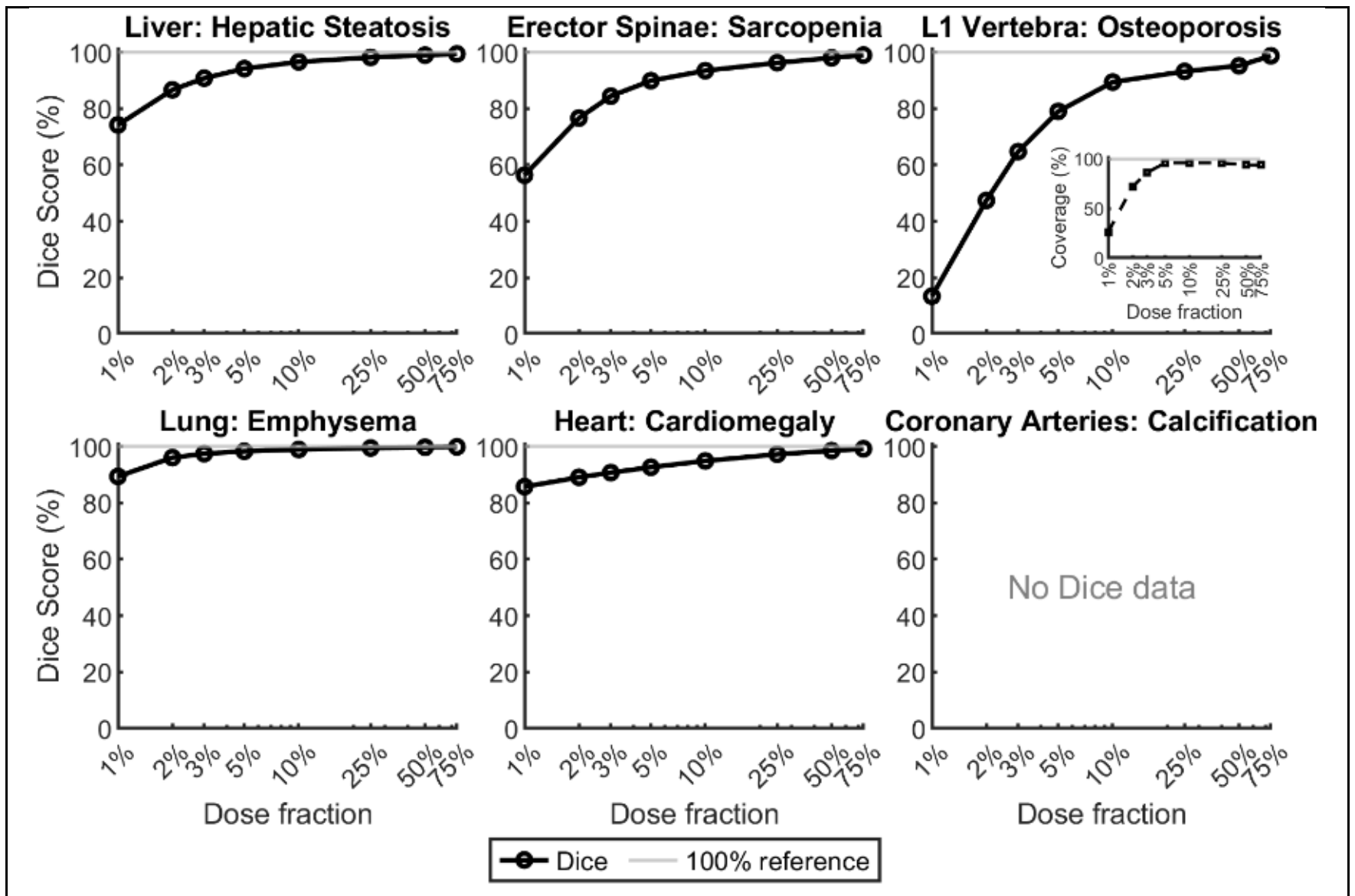

**Figure 3.** Dice similarity coefficient between low- and full-dose segmentation masks across dose fractions for evaluated organs in the Mayo low-dose CT dataset. Inset for L1 vertebra shows metric availability (coverage) across dose levels.

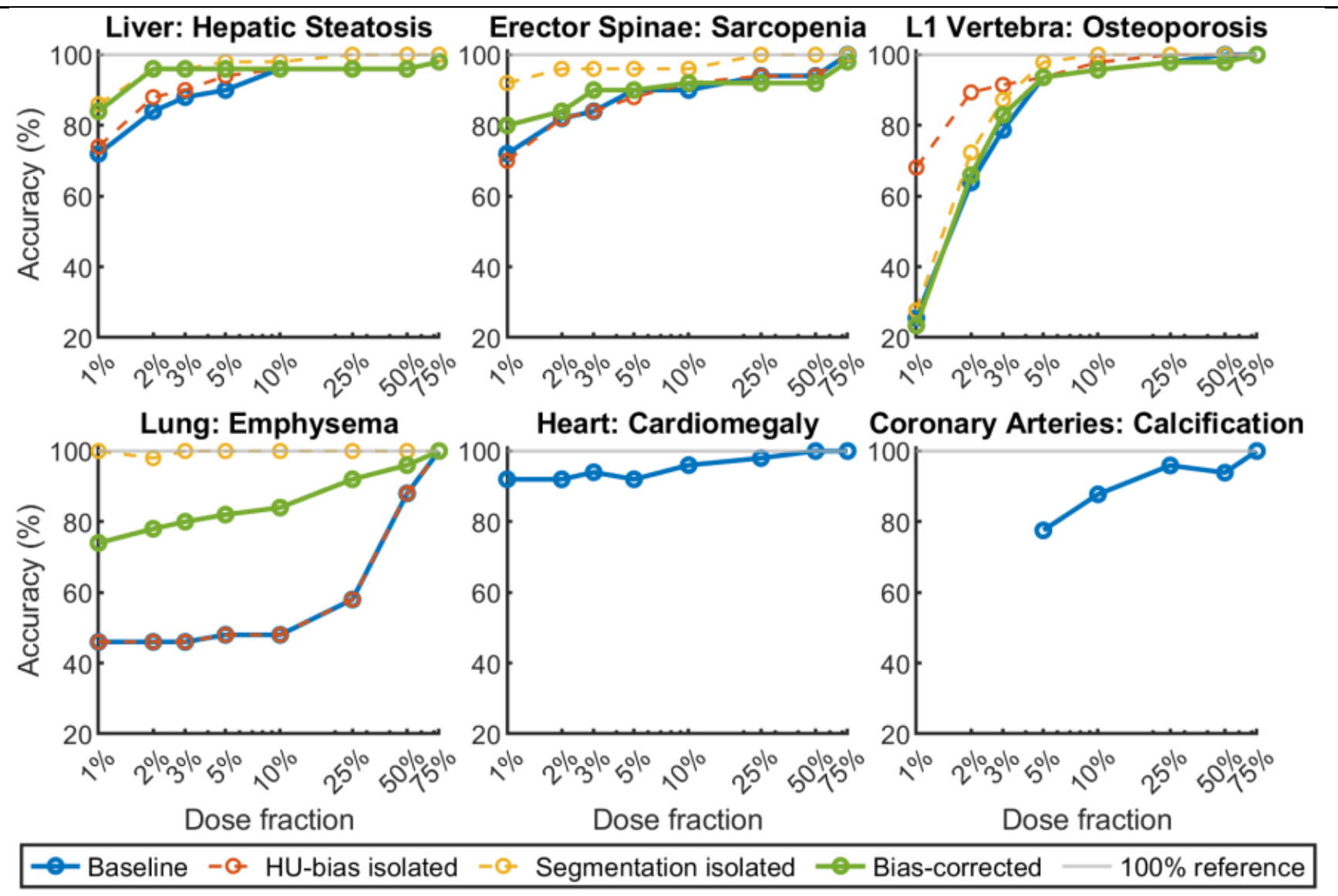

**Figure 4.** Dose-dependent classification accuracy across simulated dose fractions for each disease metric. Results are shown for the baseline (low-dose image with low-dose segmentation mask), HU-bias-isolated (low-dose image with full-dose segmentation mask), segmentation-isolated (full-dose image with low-dose segmentation mask), and bias-corrected configurations.

HU, Hounsfield unit.

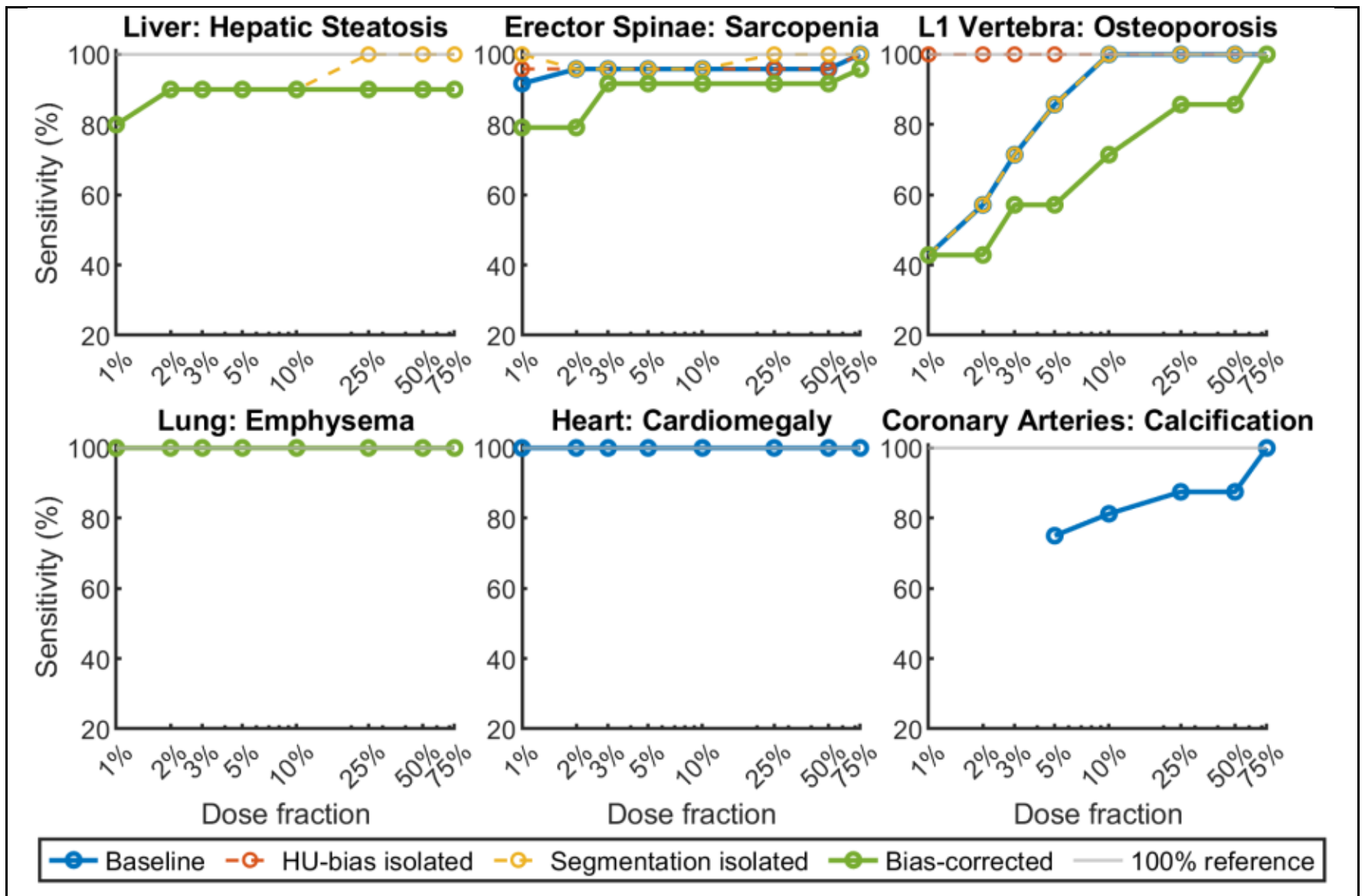

**Figure 5.** Dose-dependent classification sensitivity across simulated dose fractions for each disease metric. Results are shown for the baseline (low-dose image with low-dose segmentation mask), HU-bias-isolated (low-dose image with full-dose segmentation mask), segmentation-isolated (full-dose image with low-dose segmentation mask), and bias-corrected configurations.

HU, Hounsfield unit.

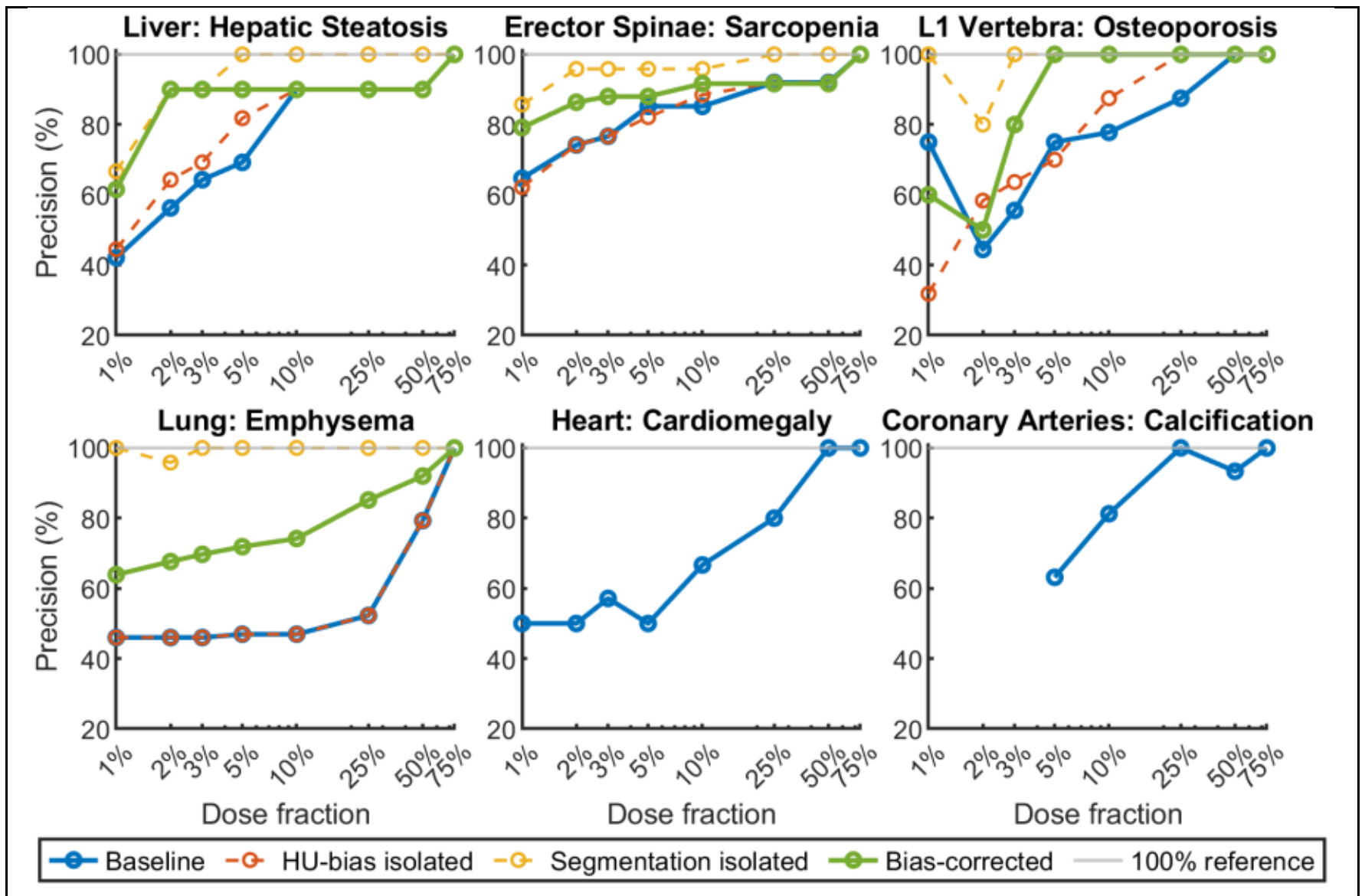

**Figure 6.** Dose-dependent classification precision across simulated dose fractions for each disease metric. Results are shown for the baseline (low-dose image with low-dose segmentation mask), HU-bias-isolated (low-dose image with full-dose segmentation mask), segmentation-isolated (full-dose image with low-dose segmentation mask), and bias-corrected configurations.

HU, Hounsfield unit.

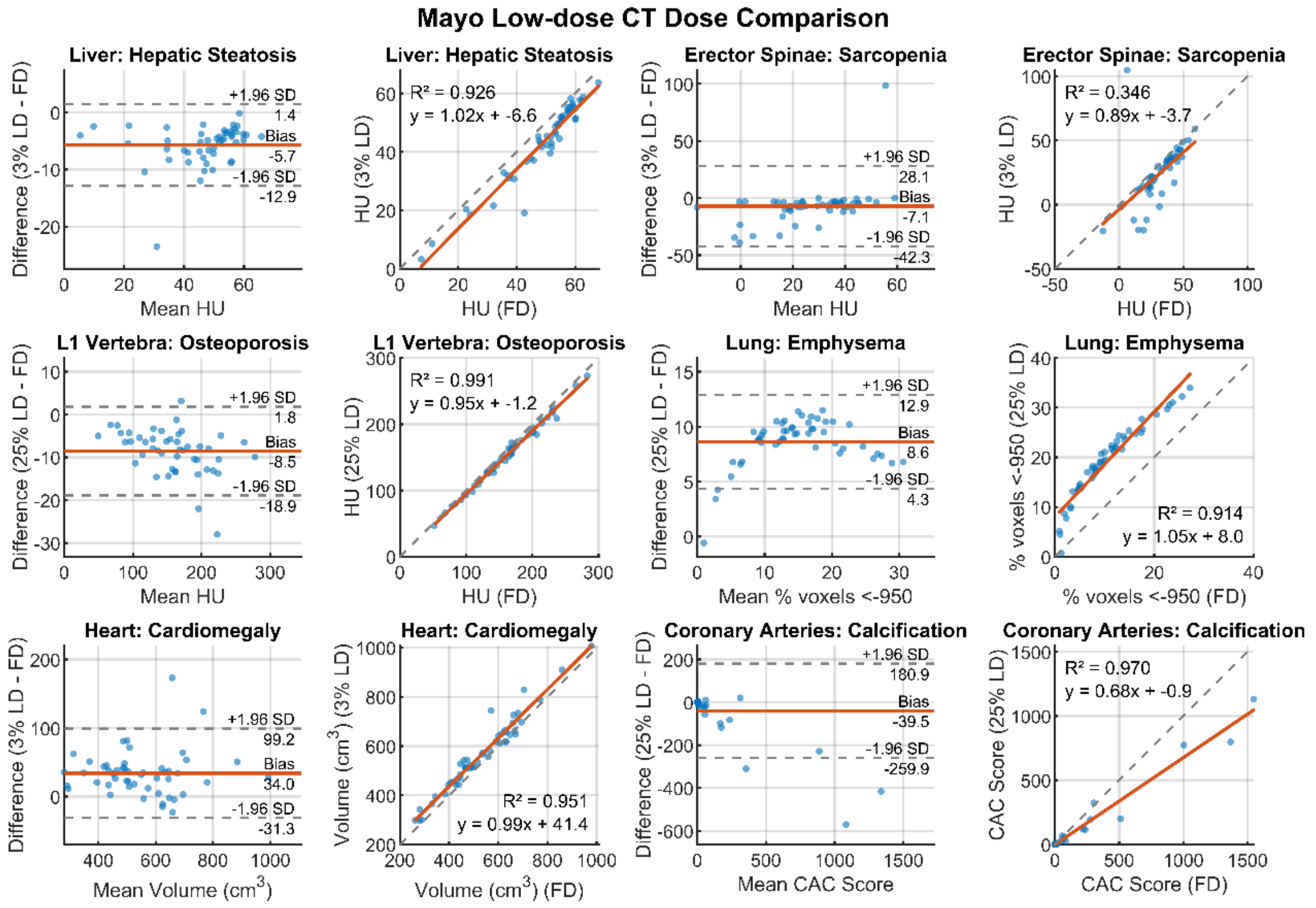

**Figure 7.** Bland-Altman plots (columns 1 and 3) and correlation plots (columns 2 and 4) comparing quantitative disease metrics between low-dose and full-dose in the Mayo LDCT dataset. For each metric, the lowest clinically relevant dose level achieving greater than 90% classification accuracy was selected (3% for chest-x-ray dose and 25% for lung cancer screening dose).

LD, low dose. FD, full dose.

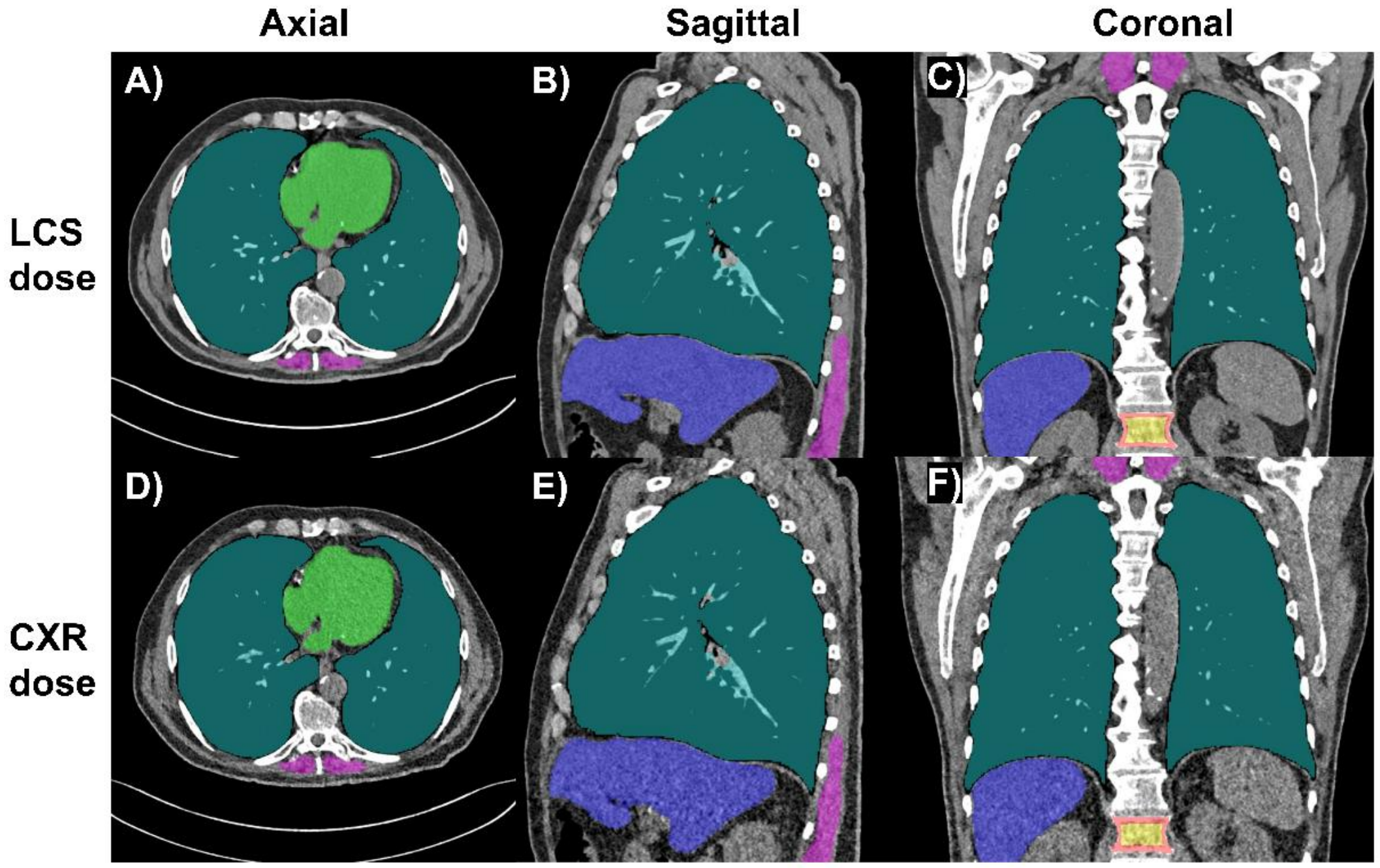

**Figure 8.** Representative axial, sagittal, and coronal views for lung cancer screening dose (LCS; A-C) and chest x-ray dose (CXR; D-F) PCCT images with overlaid organ segmentations (W/L: 400/40 HU; liver (blue), lungs (teal), erector spinae (magenta), L1 vertebra (red; yellow indicates eroded mask for trabecular region), heart (green)).

HU, Hounsfield unit.

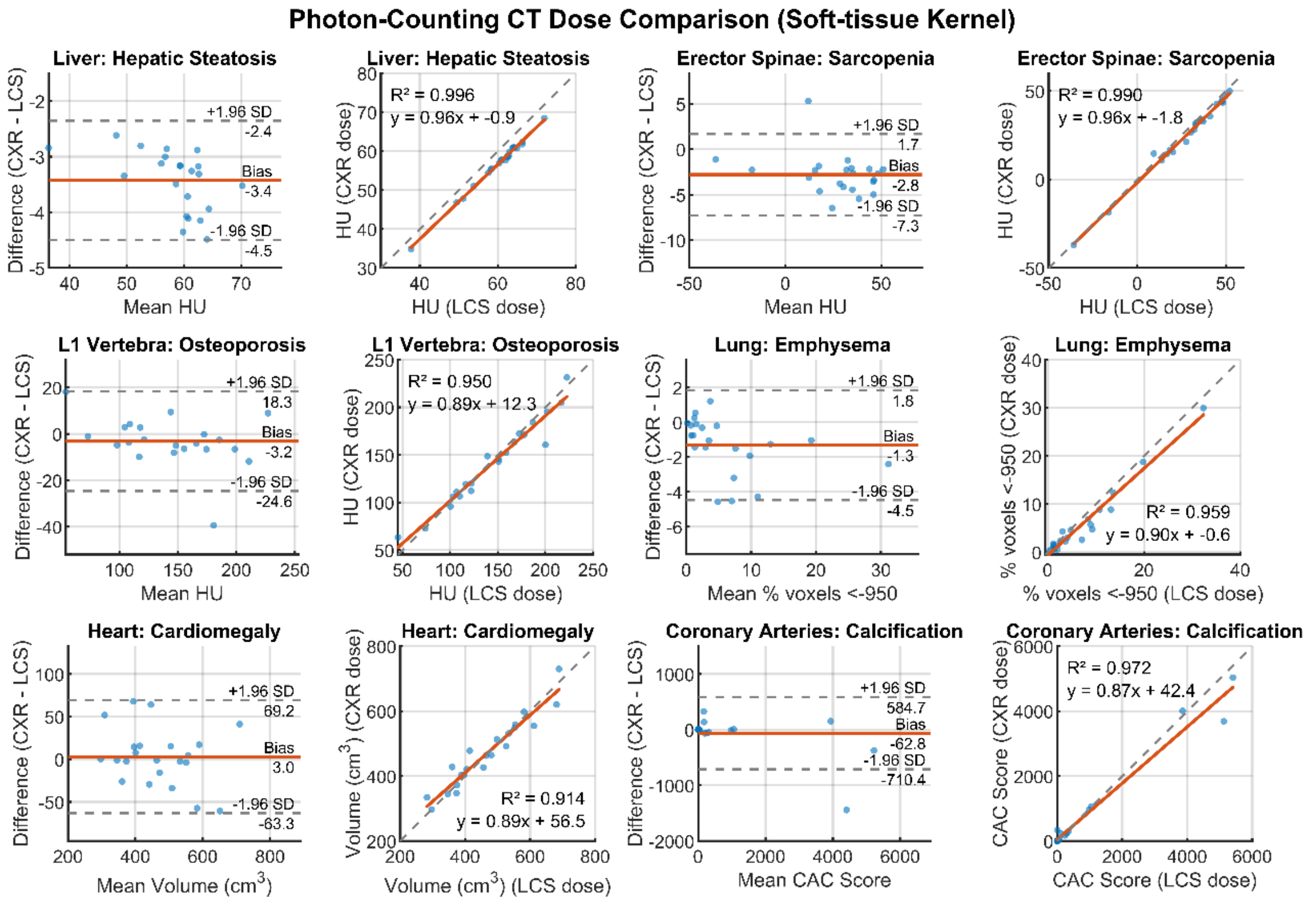

**Figure 9.** Bland-Altman plots (columns 1 and 3) and correlation plots (columns 2 and 4) comparing quantitative disease metrics between chest-x-ray (CXR) dose and lung cancer screening (LCS) dose for the photon-counting CT dataset.

**Table 1.** Summary of classification tasks, associated organs, metric types, and thresholds.

| **Organ** | **Liver** | **Erector spinae** | **L1 Vertebra** | **Heart (size)** | **Lung** | **Heart (CAC)** |
|---|---|---|---|---|---|---|
| **Condition** | Hepatic steatosis | Sarcopenia | Osteoporosis | Cardiomegaly (Enlargement) | Emphysema | Coronary atherosclerosis |
| **Metric type** | Mean HU | Mean HU | Organ interior mean HU | Volume | Volume percentage below HU threshold | CAC score |
| **Classification threshold** | < 40 HU (Zhang et al[29]) | < 30 HU (Aubrey et al[30]) | < 100 HU (Pickhardt et al[7]) | > 700 mL (Sosnowski et al[32]) | > 10% volume <-950 HU (Johannessen et al[31]) | > 10 (Hagopian et al[33]) |

HU, Hounsfield unit. CAC, coronary artery calcium